\documentclass[aps,prb,reprint,superscriptaddress]{revtex4-1}

\usepackage{graphicx}% Include figure files
\usepackage{dcolumn}% Align table columns on decimal point
\usepackage{bm}% bold math
\usepackage{setspace}

\newcommand{\fg}{f_{G}}

\newcommand{\hs}{H_{sat}}

\newcommand{\hip}{H_{||}}
\newcommand{\kx}{\kappa_x}
\newcommand{\ky}{\kappa_y}

\begin{document}

\title{Chirality-mediated bistability and strong frequency downshifting of the gyrotropic resonance of a dynamically de-stiffened magnetic vortex}

\author{Manu Sushruth}
\email{manu.sushruth@research.uwa.edu.au}
\affiliation{School of Physics and Astrophysics, M013, University of Western Australia, 35 Stirling Hwy, Crawley WA 6009, Australia.}%

\author{Jasper P.~Fried}
\affiliation{School of Physics and Astrophysics, M013, University of Western Australia, 35 Stirling Hwy, Crawley WA 6009, Australia.}%

\author{Abdelmadjid~Anane}
\affiliation{Unit\'e Mixte de Physique, CNRS, Thales, Univ.~Paris-Sud, Universit\'e Paris-Saclay, 91767 Palaiseau, France.}

\author{Stephane~Xavier}
\affiliation{Thales Research and Technology, 1 Avenue A.~Fresnel, 91767 Palaiseau, France}

\author{Cyrile~Deranlot}
\affiliation{Unit\'e Mixte de Physique, CNRS, Thales, Univ.~Paris-Sud, Universit\'e Paris-Saclay, 91767 Palaiseau, France.}

\author{Vincent~Cros}
\affiliation{Unit\'e Mixte de Physique, CNRS, Thales, Univ.~Paris-Sud, Universit\'e Paris-Saclay, 91767 Palaiseau, France.}

\author{Peter J.~Metaxas}
\email{peter.metaxas@uwa.edu.au}
\affiliation{School of Physics and Astrophysics, M013, University of Western Australia, 35 Stirling Hwy, Crawley WA 6009, Australia.}
\affiliation{School of Mechanical and Chemical Engineering, M050, University of Western Australia, 35 Stirling Hwy, Crawley WA 6009, Australia.}

\date{\today}

\begin{abstract}
We demonstrate an enhanced, bidirectional, in-plane magnetic field tuning of the gyrotropic resonance frequency of a magnetic vortex within a disk by introducing a flat edge. When the core is in its vicinity,  the flat edge locally reduces the core's directional dynamic stiffness for movement parallel to the edge. This strongly reduces the net dynamic core stiffness, leading to the gyrotropic frequency being significantly less than when the core is centered (or located near the round edge). This leads to the measurable range of gyrotropic frequencies being more than doubled and also results in a clear chirality-mediated bistability of the gyrotropic resonance frequency due to what is effectively a chirality-dependence of the core's confining potential.
\end{abstract}

\maketitle

\begin{figure}[htbp]
	\centering
	\includegraphics[width=7cm]{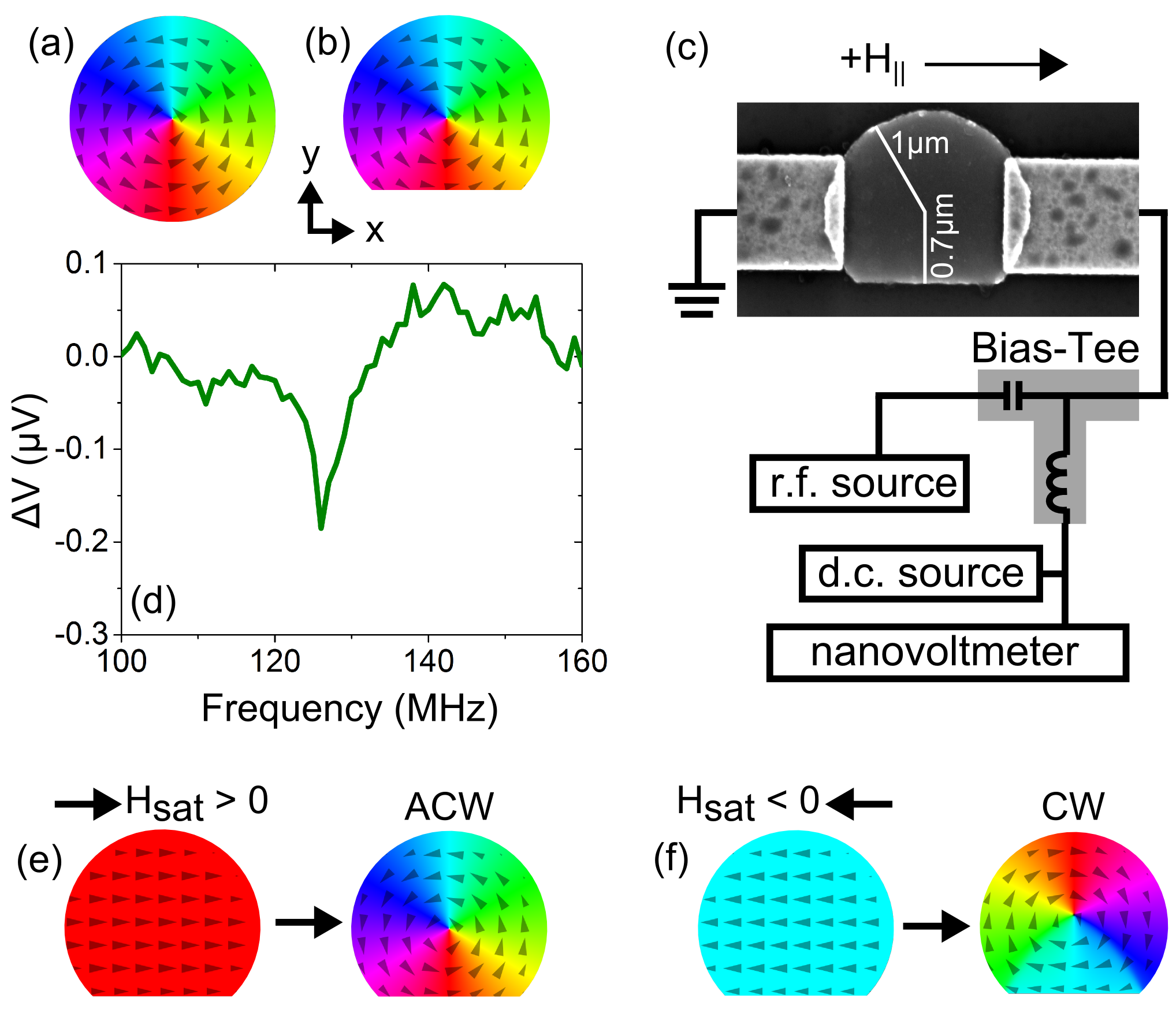}
	\caption{Simulated vortex configurations in (a) circular  and (b) chopped  NiFe-like disks in zero external magnetic field. The diameter of the circular disk is 2 $\mu$m. The geometry of the chopped disk matches that shown in (c). The (c) Schematic of the experimental setup used for probing the core dynamics including a scanning electron micrograph showing the  lateral dimensions of a chopped disk used in our experiments. (d) Experimentally obtained resonance peak at $\hip =0$. Simulated transition from the (e) right- and (f) left-oriented saturated states to, respectively, an ACW and CW vortex state ($\hip=0$). $H_{sat}$ is used experimentally to approach the saturated state and controllably produce (A) CW vortex configurations. 
	}
	\label{f1}
\end{figure}

Magnetic vortices are curled magnetization configurations that arise naturally in thin magnetic elements with lateral dimensions from $\sim 0.1$ $\mu$m to a few $\mu$m \cite{Cowburn1999-2,Shinjo2000,Wachowiak2002,Guslienko2008}. They are characterized by an in-plane, curling magnetization which surrounds an out-of-plane magnetized nano-scale region known as the vortex core [Fig.~\ref{f1}(a)]. Magnetic vortices are examples of topological solitons or defects and can display dynamic behavior which is intrinsically nonlinear \cite{Buchanan2007,Guslienko2010,Drews2012,Guslienko2008a,Lee2008a}, a characteristic which has generated significant theoretical interest. 

The potential for device applications \cite{Pigeau2010,Jung2011,Huber2011}, which include tuneable  radiofrequency signal generators/detectors \cite{Pribiag2007,Dussaux2010,Jenkins2016} and data storage devices \cite{Pigeau2010},  has strongly motivated studies in tuning the frequency of the gyrotropic resonance of magnetic vortices, $\fg$. This resonance involves the vortex core following an orbit-like path around its equilibrium position \cite{Argyle1984,Guslienko2002,Park2003,Choe2004,Novosad2005}. The gyrotropic resonance frequency is proportional to the vortex stiffness \cite{Guslienko2002}, $\kappa$, which is typically determined by a geometrically induced, primarily magnetostatic\cite{Guslienko2002,Guslienko2006a}, core confining potential (there can also be non-negligible contributions from exchange interactions\cite{Fried2016} or current-generated Oersted fields\cite{Dussaux2012}). Although the confining potential can be treated as harmonic for small radial core displacements ($\Delta E=\frac{1}{2}\kappa X^2$ where $E$ is the system energy and $X$ the radial core displacement), it is typically anharmonic\cite{Sukhostavets2013}, meaning that $\kappa$, and thus $\fg$, are both dependent on the position of the vortex core. For example, in circular\cite{Sukhostavets2013,Gangwar2015,Sushruth2016c} disks (as well as in elliptical disks, although the behavior there is slightly more complex\cite{Buchanan2006}), the gyrotropic frequency becomes higher as the core is displaced from the disk's center due to a core stiffening.

In this letter we present results on vortex core dynamics in a ferromagnetic disk that has one side which has been made flat (`chopped') [Fig.~\ref{f1}(b)]. The flat edge  enables control over the vortex chirality  \cite{Schneider2001,Huang2013b,Nakatani2005,Kimura2007,Dumas2011,Huang2010,Wu2008a} which describes the direction of the (clockwise or anti-clockwise) curling magnetization.  The chirality  is critical in this study as it determines the lateral direction that a core will be displaced under the action of  a given static in-plane magnetic field \cite{Schneider2001}. Chirality control thus enables core displacement control. Despite both the flat and round edges being repulsive to the core under static displacements towards the edge, we show that a  core undergoing gyrotropic motion near the flat edge exhibits a  strongly reduced  \textit{dynamic} stiffness when moving along the flat edge (i.e.~perpendicular to the direction of the static displacement). As a result, the net dynamic core stiffness \cite{Buchanan2006} is also reduced, leading to  $\fg$ at the flat edge being lower than $\fg$ for the disk-centered core. $\fg$ can thus either be strongly increased (by shifting the core towards the disk's round edge) or decreased (by moving the core towards the flat edge). This leads to the  range of accessible $\fg$-values being approximately doubled (i.e.~we can both significantly increase and decrease $\fg$ relative to its value for a non-displaced vortex in zero field). We note that frequency downshifting has also been  observed in square ferromagnetic elements\cite{Langner2013,Cui2016} and, to a lesser degree, in triangular elements \cite{Yakata2013}  when moving the core towards a flat edge within the element.  

A final consequence of the geometrical asymmetry in our system is that the vortex is characterized by a clear chirality-induced, dynamic bistability. Indeed, for a given, finite, static in-plane magnetic field (applied such that it acts to displace the core perpendicular to the disk's flat edge),  two values of $\fg$ can be observed depending on the vortex chirality.  This is reminiscent of the {polarity}-induced bistability studied by de Loubens \textit{et al} \cite{deLoubens2009}. Here however, it is chirality-mediated and arises because, for a given in-plane field polarity, the chirality determines which part of the asymmetric confining potential the gyrating core is subject to.

A scanning electron microscope image of a chopped disk is shown in Fig.~\ref{f1}(c). The distance from the disk's center to the round edge is 1 $\mu$m (radius of the circular part of the disk) while the shortest distance from the disk's center to the flat edge is 0.7 $\mu$m. Disks were fabricated from a continuous sputtered //NiFe(30 nm)/Au(8 nm) layer via Argon ion milling using a hard Ti mask defined by electron beam lithography (NanoBeam Limited). Evaporated lateral contacts were defined using electron beam lithography and liftoff. Auger spectroscopy suggests that approximately 30 nm of Ti remains on the top of the NiFe/Au chopped disk \cite{Sushruth2016c}.  The measured device was wire bonded to a sample mount which was placed between the poles of an electromagnet in such a way that a static, in-plane magnetic field, $\hip$, could be applied along the flat edge of the disk (note that simulations detailed below will suggest the presence of a small misalignment  of $\sim 3^{\circ}$ between the field and the disk's flat edge). A nanovoltmeter (200 ms integration time) was used to measure the voltage across the device in the presence of injected dc  or rf currents [Fig.~\ref{f1}(c)]. The rf source output power was -14 dBm.

An  rf current  injected across the device generates a transverse, rf Oersted field in the lower NiFe layer which can drive gyrotropic core motion\cite{Sushruth2016c}. On resonance, this  generates an oscillation in the sample's resistance (via anisotropic magnetoresistance, AMR) which can mix with the input rf current to generate a measurable rectified voltage and enable the electrical identification of $\fg$\cite{Kasai2006,Kim2010b,Goto2011,Goto2011a,Gangwar2015,Sushruth2016c}. In Fig.~\ref{f1}(d) we show the rectification peak obtained in the chopped disk at zero applied field from which we can extract $\fg\approx 126$ MHz. This value matches closely with the simulated $\fg=128$ MHz at $\hip =0$ (see below). Note that in circular disks, a rectification peak is not generated for $\hip = 0$ due to the symmetry of the core trajectory around the center of the disk which results in the generated voltage time-averaging to zero\cite{Sugimoto2014,Sushruth2016c}. In contrast, the chopped disk geometry leads to the core's equilibrium position being shifted away from the circular center of the element even for $\hip =0$ [Fig.~\ref{f1}(b)] which enables the generation of a finite amplitude peak. This result is discussed further in Supplementary Figure 1\footnote{Supplementary information to be provided at a later date.}.

\begin{figure}[htbp]
	\centering
	\includegraphics[width=6cm]{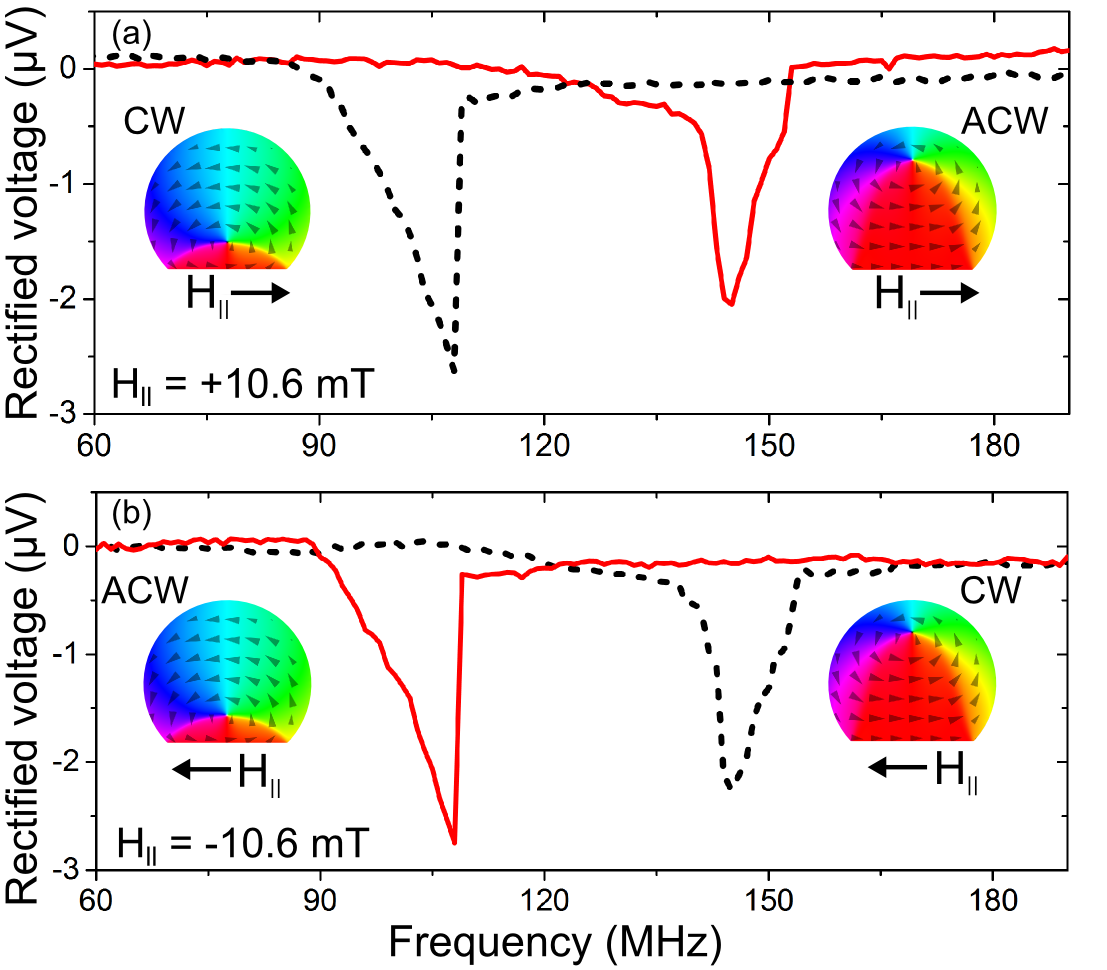}
	\caption{Experimentally obtained magnetoresistive rectification traces demonstrating chiral bistability in a chopped disk at $\mu_0 \hip$ values of (a) $+10.6$  mT and (b) $-10.6$mT for CW and ACW vortices. Insets show simulated at-equilibrium vortex configurations with shifted cores ($\mu_0 \hip =\pm 10$ mT). For simplicity, each peak has been given a negative sign (the peak sign will depend on the core polarity\cite{Goto2011,Sushruth2016c} which was not controlled in these experiments). }
	\label{f4}
\end{figure}

The vortex chirality defines the direction of the core shift for a given $\hip$\cite{Schneider2001}. For example, a positive $\hip$ shifts the core to the round edge for an ACW vortex but shifts it to the flat edge for a CW vortex [insets of Fig.~\ref{f4}]. 
Micromagnetic simulations performed using the MuMax3 micromagnetic code\cite{Vansteenkiste2014} demonstrate the ability to choose the vortex chirality in the chopped disk by starting from an in-plane saturated state and then returning the field to zero [Figs.~\ref{f1}(e,f)]. In the simulations, we used 30 nm thick cells with lateral dimensions of  $\sim 3.9 \times 3.9$ nm$^2$ ($512\times 512$ cells for the $2\times 2$ $\mu$m$^2$ simulation region)\footnote{Frequencies obtained using  a full z-discretization (8 cells) were within 
1.6 MHz of those obtained with a single z-discretization.}. We then  initialized the system with a uniform magnetization parallel to the flat edge and let the magnetization evolve towards a final, converged state in zero magnetic field with damping parameter $\alpha =0.01$ (saturation magnetization = 800 kA/m; exchange stiffness = 13 pJ/m; nil anisotropy; $\gamma=1.7595\times 10^{11}$ rad(T.s)$^{-1}$ which is within $\sim 5$ \% of the experimentally determined value \cite{Shaw2013}).  We find that  the initial direction of the magnetization along the flat edge  remains fixed upon the return to zero field and thus  defines the chirality of the relaxed, curled vortex state [Figs.~\ref{f1}(e,f)]. Note that the simulated transition to the single vortex state proceeds via a non-trivial intermediate state analogous to that seen in Co disks with widths $>800$ nm\cite{Dumas2011}.  Experimentally,  the left and right saturated magnetic states are approximated by applying an in-plane saturating field, $\hs=\pm 0.6$ T,   along the flat edge of the disk [Figs.~\ref{f1}(e,f)]. In Supplementary Figure 2\cite{Note1} we confirm our ability to choose the core chirality by exploiting differences in the vortex annihilation fields at the flat and round edges\cite{Huang2013b} (as well as general differences in the quasi-static magnetization reversal process). Dynamic results presented below will also confirm  our ability to set the chirality and controllably displace the core towards the flat or round edge of the disk.

We now use the on-resonance rectification to experimentally measure the influence of the core's equilibrium position on $\fg$. We first set a CW vortex chirality, return to zero field and then apply $\mu_0\hip=+10.6$ mT. This shifts the core towards the disk's flat edge  [Fig.~\ref{f4}(a), left]. An rf signal  is then injected across the disk, sweeping from low frequency to high frequency, enabling us to determine $\fg= 108$ MHz  (black dashed lines in Fig.~\ref{f4}(a)). Note that the measured frequency is significantly lower than the value found at  $\hip =0$ in Fig.~\ref{f1}(d) ($\approx 126$ MHz). 
Repeating the measurement for an ACW chirality at the same field allows us to probe core dynamics at the round edge of the disk  [Fig.~\ref{f4}(a), right] for which we find $\fg =140$ MHz   [solid red line in Fig.~\ref{f4}(a)]. This confirms the chiral bistability in that  we can observe two different values of $\fg$ for a fixed $\hip$, depending on the vortex chirality. 
Changing the polarity of  $\hip$ changes the direction of the core displacement for a given chirality. As such, for $\mu_0\hip=-10.6$ mT [Fig.~\ref{f4}(b)], it is now the CW vortex core which moves to the round edge of the disk, resulting in it having the higher $\fg$.

\begin{figure}[htbp]
	\centering
	\includegraphics[width=7cm]{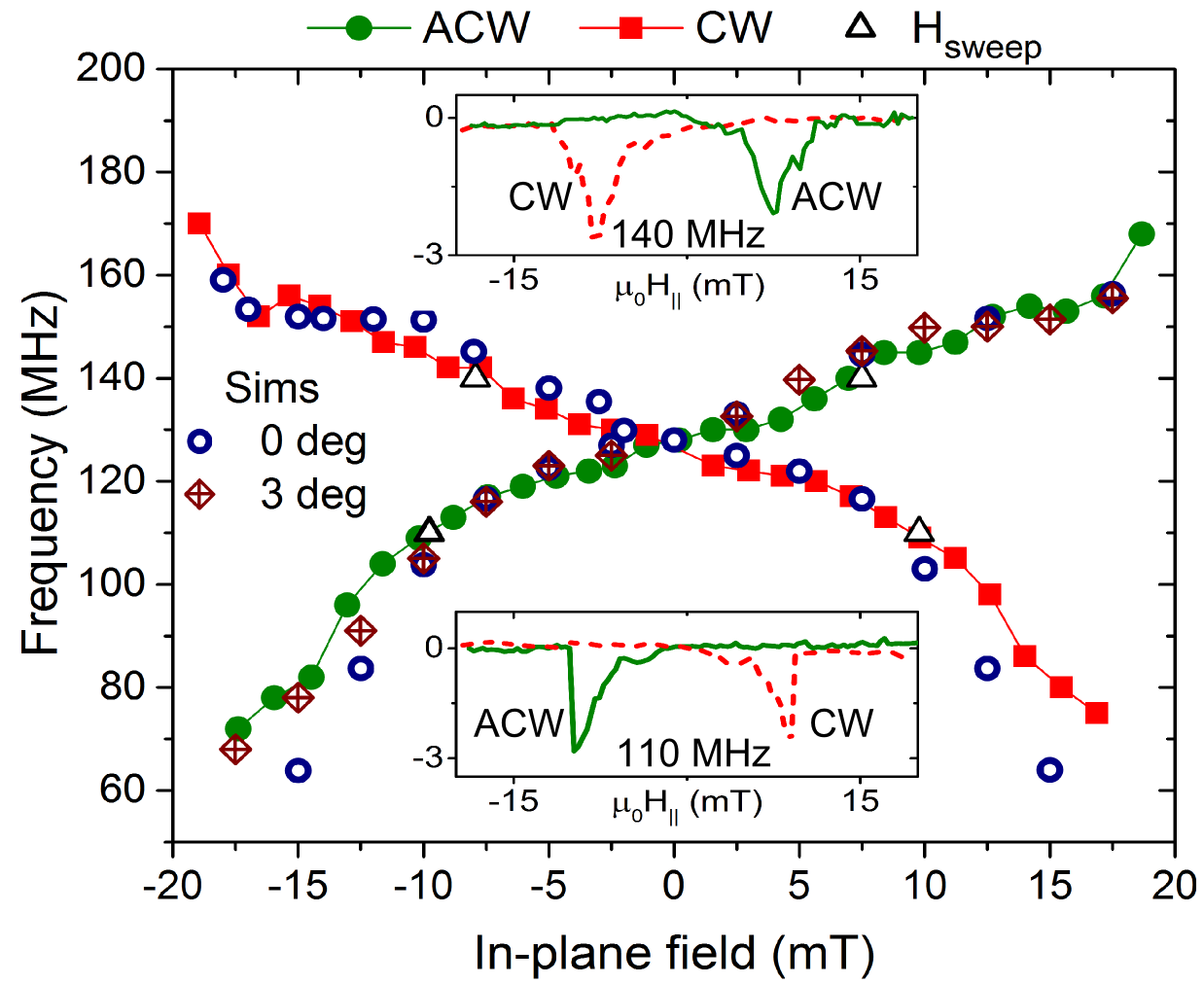}
	\caption{Experimental data showing the gyrotropic resonance frequency as a function of $\hip$ for CW (solid squares) and ACW (solid circles) vortices. Four data points obtained from field-swept measurements (see insets) at 140 Mhz and 110 MHz  are plotted as open triangles. The plot also shows simulated gyrotropic frequencies for a perfect alignment of the flat edge and $\hip$ (`0 deg'; shown for both chiralities) and for a misalignment of 3$^{\circ}$ (`3 deg'; shown for ACW).}
	\label{f5}
\end{figure}

The results of experiments carried out for both chiralities are given in Fig.~\ref{f5}, showing the full evolution of $\fg$ with $\hip$. An example of peaks obtained  by carrying out field sweeps at fixed frequency (field-resolved measurements) are also shown as insets in Fig.~\ref{f5} and show good agreement with the frequency-resolved data.  Note that there are two data sets in Fig.~\ref{f5} corresponding to CW (filled squares) and ACW (filled circles) vortices.  Here, it is again clear that for a given finite in-plane field, $\fg$ can be either above or below the zero field value of $\fg$ depending on the vortex chirality with low frequencies obtained when the core is shifted towards the flat edge ($\hip <0$ for an ACW vortex and $\hip >0$ for a CW vortex). Also note that the difference between $\fg$ values for the two chiralities  increases with $\hip$ as the core is pushed further away from the disk's center.  Simulated gyrotropic frequencies (open circles) for different chiralities and $\hip$ values  are also shown in Fig.~\ref{f5} with good agreement between the simulation and experimental results. Note however that  the best agreement is obtained by including a misalignment of $3^{\circ}$ between $\hip$ and the flat edge (this causes the core to be slightly displaced towards the point where the flat edge meets the disk's round edge, slightly increasing the stiffness and thus $\fg$; see Supplementary Figure 3).  To obtain simulated $\fg$ values, we performed field-pulse-driven `ringdown' simulations\footnote{After relaxing the system at the given $\hip$ value [applied along the $x$-axis as per Fig.~\ref{f1}(a)], we apply an in-plane sinc field pulse along the $y$-axis [as per Fig.~\ref{f1}(a)] with an amplitude of 2 mT, a 300 ps time offset and a cut-off frequency of 30 GHz. This induces damped gyrotropic core dynamics around the core's $\hip$-dependent, equilibrium location.  The resulting time trace of the $x-$component of the system's spatially averaged magnetization is then Fourier analyzed to extract $\fg$.} (e.g.\cite{McMichael2005,Fried2016a}) for a number of $\hip$ values. 

To understand the observed drop-off in $\fg$ for cores at the flat edge of the disk,  we look at the relationship  between $\fg$ and the core stiffness which depends not only on the location of the core but the direction of core movement\cite{Buchanan2006}:
\begin{equation}
\fg = \frac{ \sqrt{\kappa_{x}\kappa_{y}}}{2\pi G}.
\label{efg}
\end{equation}
Above, $\kx$ and $\ky$ correspond to the local core stiffness in the $x$ and $y$ directions [defined in Fig.~\ref{f1}(a,b)] and $G$ is the gyroconstant \cite{Thiele1973,Huber1982,Guslienko2002} (considered field-independent here). Note that in a circularly symmetric system\cite{Guslienko2006a,Guslienko2002}, $\fg$ is simply given by $\kappa/2\pi G$ with $\sqrt{\kx\ky}$ here replacing the otherwise direction-independent $\kappa\equiv$``$\kappa_{tot}$''. The stiffness coefficient along an axis $x$ is defined via $\Delta E=\frac{1}{2}\kappa_xx^2$ (valid for small\cite{Sukhostavets2013} $x$) where $E$ is the energy of the system. An accurate description of dynamics must however take into account the energy of the moving core rather than static displacements\cite{Buchanan2006,Fried2016}. To extract the dynamic $\kappa$ from micromagnetic simulations, core dynamics were driven using an in-plane sinusoidal field with an amplitude of 5 $\mu$T and frequency equal to $\fg$ as obtained from the ringdown simulations\cite{Fried2016a}. $\kx$ and $\ky$ were then extracted from parabolic fits to the total system energy as a function of core displacement along the $x$ and $y$ directions respectively.

\begin{figure}[htbp]
	\centering
	\includegraphics[width=8cm]{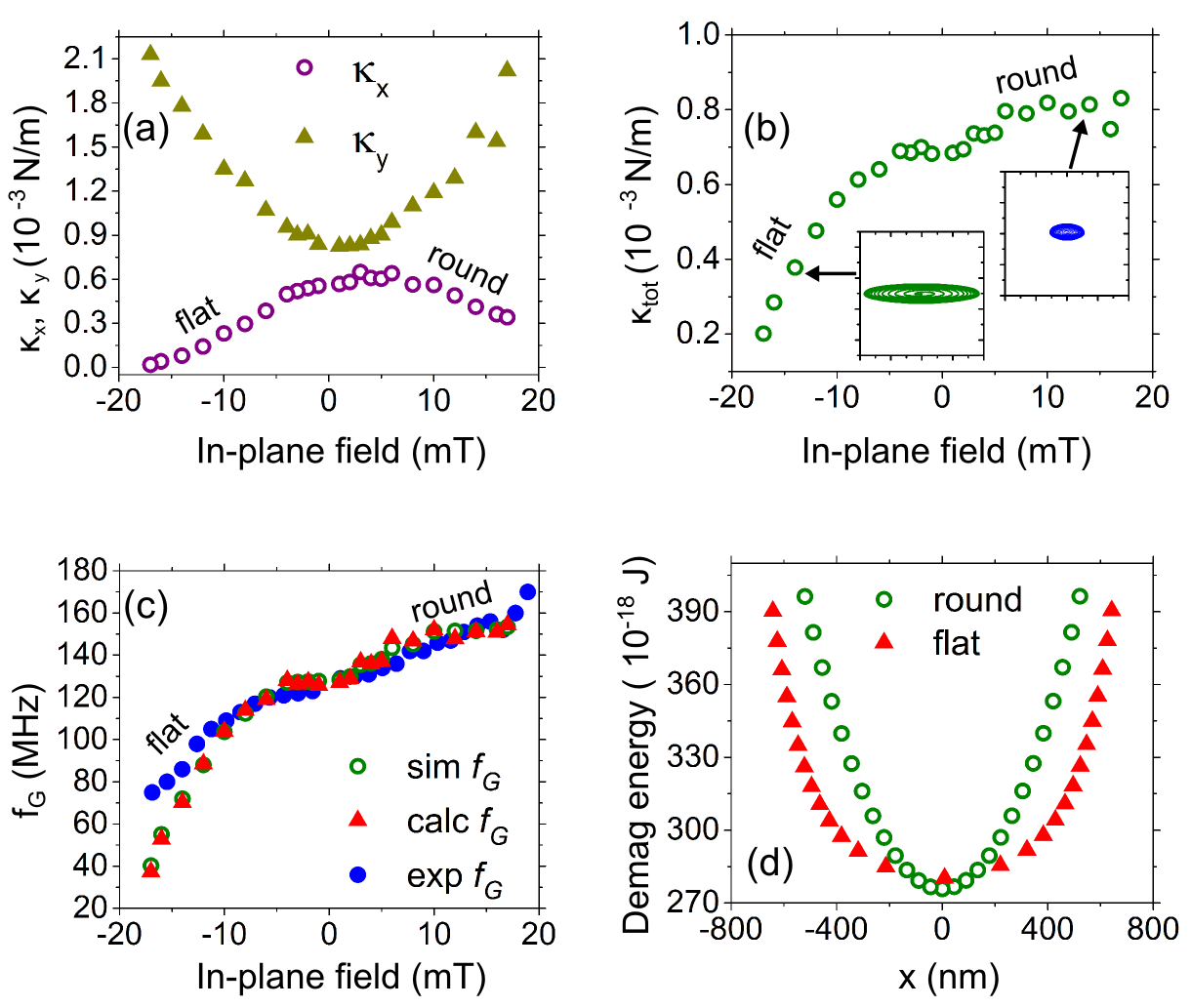}
	\caption{(a)  $\kappa_{x}$ and $\kappa_{y}$ and (b) $\kappa_{tot}=\kappa_x\kappa_y$ as extracted from simulations  for an ACW vortex as a function of $\hip$.  (c) Simulated, calculated [via Eq.~(\ref{efg})] and experimental values of $\fg$ as a function of $\hip$ for the ACW vortex. (d) Demagnetizing energy profile extracted from simulation for a core displacing approximately along the flat edge (solid triangles) and round edge (open circles). The energy is shown  as a function of equilibrium core displacement along the $x$-direction (induced by application of a stepped field along the $y$-direction; i.e. perpendicular to the flat edge).  Insets in (b) show the core trajectory at resonance when the core is displaced towards the flat edge ($\mu_0\hip=-15$ mT) or towards the round edge ($\mu_0\hip=+15$ mT).}
	\label{f3}
\end{figure}

Fig.~\ref{f3}(a) shows the extracted $\kx$ and $\ky$ values versus in-plane field for an ACW vortex for which a  negative $\hip$ will displace the core toward the disk's flat edge. Although  $\ky$ is relatively symmetric about $\mu_0\hip=0$ mT, $\kx$, related to displacements along the flat edge of the disk,  is clearly asymmetric and lowest when the core is located near the flat edge ($\hip <0$). This is despite that edge of the disk being repulsive in terms of static displacements of the core towards it [i.e.~the equilibrium position of the static core at $\hip =0$ remains relatively close to the disk's center; Fig.~\ref{f1}(b)]. The asymmetry in $\kx$ is critical to understand the drop-off in $\fg$ as it leads  to a strong local reduction in the $\sqrt{\kx\ky}\equiv \kappa_{tot}$ term [Fig.~\ref{f3}(b)] which in turn leads to a strong reduction in $\fg$ as per Eq.~(\ref{efg}). Indeed the $\fg$ predicted from the extracted $\kx$ and $\ky$ values using  Eq.~(\ref{efg}) can be used to accurately reproduce the simulated frequencies which in turn approximate the measured frequencies well [Fig.~\ref{f3}(c)]. 

The reduced value of $\kx$ at the disk's flat edge is primarily due to a weaker dependence of the demagnetizing-energy on the $x$-position of the core at that side of the element (contributions from demagnetizing, exchange and Zeeman energies are compared in Supplementary Figure 4\cite{Note1} for the chopped and unchopped disk). This weaker dependence can be directly visualized in Fig.~\ref{f3}(d) where we show the change in the system's demagnetizing energy for a core that is quasi-statically shifted approximately laterally along the flat and round edges via a $y$-oriented magnetic field. One sees that small lateral displacements near the curved edge lead to larger changes in the demagnetizing energy (per unit displacement) than do lateral displacements along the flat edge. At least for static magnetization configurations with laterally shifted cores, the demagnetizing energy (as well as the displacement-induced change in the demagnetizing energy) is concentrated within lines that join the core to positions which are close to the meeting points of the flat and curved edges of the disk (these lines are essentially domain walls; see Supplementary Figure 5\cite{Note1}).  As an aside, we note that the reduced $\kx$ at the flat edge also has a strong effect on the orbit, with simulations predicting that the orbit will be strongly elongated in the $x$-direction (lower inset in Fig.~\ref{f3}(b)) as compared to the orbit near the curved edge (upper inset in Fig.~\ref{f3}(b)).

In conclusion, we have shown that the introduction of a flat edge into a magnetic disk can generate  a chirality-driven bistability of the vortex gyrotropic resonance frequency as well as an increased rage of accessible gyrotropic resonance frequencies. Indeed, depending on the vortex chirality, a static in-plane magnetic field can drive the core either towards the disk's round edge (where the gyrotropic frequency is known to increase) or towards its flat edge (which is shown here to reduce the gyrotropic frequency). The latter frequency downshifting is shown to be a dynamic effect related to the core moving both parallel and perpendicularly to the flat edge during gyrotropic motion. Calculations of the stiffness of the resonating core demonstrate that the core's dynamic stiffness  along the flat edge ($\kappa_x$) is strongly reduced when the core is close to the flat edge, resulting in a lower net dynamic stiffness ($\propto \sqrt{\kappa_x}$) and  lower gyrotropic resonance frequencies.

\acknowledgments
This research was supported by the Australian Research Council's Discovery Early Career Researcher Award scheme (DE120100155), a research grant from the United States Air Force (Asian Office of Aerospace Research and Development, AOARD) and the University of Western Australia's Early Career Researcher Fellowship Support scheme. The authors acknowledge resources provided by the Pawsey Supercomputing  Centre with funding from the Australian Government and the Government of Western Australia as well as the facilities, and the scientific and technical assistance of the Australian Microscopy \& Microanalysis Research Facility at the Centre for Microscopy, Characterisation \& Analysis, The University of Western Australia, a facility funded by the University, State and Commonwealth Governments. This work was performed in part at the WA node of the Australian National Fabrication Facility, a company established under the National Collaborative Research Infrastructure Strategy to provide nano and micro-fabrication facilities for Australia's researchers. Thanks to M.~Kostylev for useful discussions.

\end{document}